# New Optimized Band-Pass Filter To Increase Optical Temporal Coherence of Thermal Light

**Anatoliy I Fisenko, Vladimir F Lemberg**

*ONCFEC Inc., 250 Lake Street, Suite 909, St. Catharines, Ontario L2R 5Z4, Canada*

*Phone: 905-931-9097, E-mail: afisenko@oncfec.com*

## Abstract

A new optimized band-pass filter for wavelengths is proposed to increase the optical temporal coherence of thermal light. The choice of parameters for this filter is based on solving an optimization problem for finding the most intensely emitted frequency interval in the black-body radiation spectrum. The calculated frequency interval is also the narrowest band for the coherence filter for a given value of the total transmitted energy. As a result, the interval found can be used to achieve the highest possible coherent properties at a given level of total energy that has passed. The achieved coherence length values for such optimized band-pass filters are calculated. Analytical results are given for finding optimal intervals and calculating the coherence of optimized filters.

# Introduction

The optical coherence of thermal light sources has been the subject of growing interest in recent decades. Contrary to popular belief that heat light is not fully coherent, as was shown in [1], heat radiation has a finite coherence length.

The knowledge of the coherence length of the thermal light is very important for the design of various photonic devices [2].

For example, the knowledge of the coherence length is very important for the design of the solar micro-antennas and rectennas [3-5]. The maximal size of the single micro-antenna cell is limited by the coherence length of the incident thermal light [6, 7].

The magnitude of the coherence length is also important for the design of traditional photovoltaic devices that are based on the use of semiconductor p-n-junctions. The multilayered design of such devices allows to increase the absorption of incident sunlight [8-11]. For better absorption, the thickness of a single layer is also limited by the coherence length of the light [12–15].

The same approach applies to the nano-photonic materials. The size of a single cavity should be limited by the value of the coherence length to provide the better absorption [16-19].
Partially coherent light is also used for the numerous applications using the Fabry-Perot interferometry [20, 21]. In this case the coherence length of the light is also limits the maximum size of the measured irregularities, such as micro-cavities, pores, etc.

The partially coherent sources of thermal light are also used in a thermo-light holography (thermal ghost imaging) [22-24]. In this technology, partially coherent thermal light interferes with the light from completely coherent sources. The area of the produced by the interference image is limited by the coherence length of thermal light.

The coherence of thermal sources is also important in other areas. In [25] it was proposed to use the thermal emitters for stealth communication in the terahertz frequency range.

The degree of the coherence of thermal light can also affect the rate of photo-chemical reactions. The effect of the coherence of heat light on photosynthesis and growth plans in agriculture was also studied in [26].

In all the above cases, the degree of coherence of thermal light is a key factor in the design of various devices. Thus, the search for sources of thermal radiation capable of generating coherent emission is an important task for many applications.

The most natural way to increase the coherence of thermal light is to use narrow-band frequency filters. At the same time, narrow-band filtering reduces the amount of the total emitted energy. Thus, the choice of the filter parameter represents a significant trade-off between the achieved level of coherence and the reduced value of total radiation.

In this article, we propose a new optimized band-pass filter to increase the coherence of thermal light. This filter provides the best values of the total emitted radiation at a given width of the frequency interval. The parameters of this new filter are selected in accordance with the solution of the optimization task for finding the most intensively emitting interval of the blackbody spectrum of thermal radiation [27]. It is expected that this filter will provide the best coherent properties at a given value for the total intensity of the transmitted light.

In Chapter 1, we describe the idea of an optimized filter to increase the coherence of thermal light. The rules for choosing the parameters for this filter are described, and optimization tasks are formulated for this.

Chapter 2 proposes a solution for this optimization task. For an ideal filter, this solution is associated with finding the most intensely emitted interval in the spectrum of thermal radiation, solved in our previous publication [27]. Here we use the previously obtained results to select the parameters of an optimized coherent filter.

In Chapter 3, we investigate the coherent properties of the optimized filter. Specifically, we calculate the coherence length of the optimized filter for different values of the interval $\Delta$. We also found a connection between the obtained coherence length and the received amount of the total energy of the transmitted light. The results show how you can achieve a significant increase in the length of the correlation with the minimum possible energy loss. The obtained results can be used to design and optimize various devices.

Chapter 4 presents analytical results for the calculated coherence function of thermal light transmitted through an optimized coherent filter.

In Chapter 5, we discuss the possible practical applications for the obtained results.

# 1. Idea of the Optimized Coherent Filter

The idea of the proposed coherent filter is to select the interval of frequencies $[v_1, v_1+\Delta]$ (here $v_1$ is a lower frequency, $\Delta$ is width of the interval) in such an optimized way that the total energy $E_p(v_1, v_1+\Delta)$ of the transmitted thermal light reaches the maximal possible value for a given width of the interval $\Delta$:

$$E_p(v_1, v_1+\Delta) = \max; \ (\Delta = \text{const}) . \tag{1.1}$$

The interval $[v_1, v_1+\Delta]$, which satisfies to the condition (1.1), is preferable use as an ideal coherent filter for the following reasons:
   a) this interval represents the most intensively radiating interval in the spectrum. Thus, the choice of this interval for the filter provides the minimum possible loss of the emitted energy.
   b) reverting the statement (a), this interval also represents a narrowest possible range of frequencies for a given value of the total radiation energy $E_p(v_1, v_1+\Delta)$. It can be expected that the narrowest range of frequencies will give the best (or very close to the best) coherent properties of the transmitted light.

Thus, these reasons allows us to use the optimization task (1.1) as a criterion for creating the optimal filter for increasing the coherent properties of thermal light.

The parameters of this filter (in particular, the range of frequencies $[v_1, v_1+\Delta]$ can be found by solving this optimization problem (1.1).

# 2. Optimization Task: Search for Optimized Coherent Filter Parameters

The total energy of the passed through the filter thermal light $\underline{E}_p(v_1, v_1+\Delta)$ in the optimization task (1.1) can be expressed in the most common form:

$$E_p(v_1, v_1+\Delta) = \int_{v_1}^{v_2} \alpha(v)\varepsilon(v,T)I_{BB}(v,T)dv , \tag{2.1}$$

where $0 \leq \alpha(v) \leq 1$ is the filter transfer function, $0 \leq \varepsilon(v,T) \leq 1$ is the spectral emissivity of the emitter, T is the temperature of the emitter, and $I_{BB}(v,T)$ is the radiant spectral energy density of black-body radiation

$$I_{BB}(v,T) = \frac{8\pi h}{c^3} \frac{v^3}{e^x - 1}, \qquad (2.2)$$

where $x = \frac{hv}{k_B T}$ is the arbitrary parameter, $h$ is the Planck constant, and $k_B$ is the Boltzmann constant.

In the case of the ideal filter

$$\alpha(v) = 0; \text{ (for } v < v_1) \qquad (2.3)$$
$$\alpha(v) = \text{const}; \text{ (for } v_1 < v < v_2)$$
$$\alpha(v) = 0; \text{ (for } v > v_2).$$

and for "grey-body" thermal emitter

$$\varepsilon(v,T) = \varepsilon_0 = \text{const}, \qquad (2.4)$$

the optimization problem (1.1) is reduced to the search for the most intensely emitting frequency interval in the black-body radiation spectrum. This optimization task was solved in [27]. The interval $[v_1, v_1+\Delta]$ with a maximal value of the total energy density $E_p (v, T)$ can be found from the following conditions:

$$\frac{\partial E_p(v_1, v_1 + \Delta, T)}{\partial v_1} = 0, \qquad \frac{\partial^2 E_p(v_1, v_1 + \Delta, T)}{\partial v_1^2} < 0. \qquad (2.5)$$

As it was shown in [27, see the chapters 5.1, 5.2], using the polylogarithmic representation for the function $E_p(v_1, v_1 + \Delta, T)$ [28, 29], the conditions (2.5) can be presented in the form:

$$x_1^3[\exp(x_1 + s) - 1] - (x_1 + s)^3[\exp(x_1) - 1] = 0 \qquad (2.6)$$

or, equivalently

$$I_{BB}(x_1, T) = I_{BB}(x_1 + s, T) = I, \qquad (2.7)$$

where $x_1 = \frac{hv_1}{k_B T}$ is the arbitrary parameter for the beginning of the interval, and $s = \frac{\Delta}{k_B T}$ is the width of the interval in arbitrary units. Thus, the optimal (in terms of the total energy $E(v, T)$) interval can be found from the intersections of the Planck distribution curve with the parallel lines $I = $ constant. Following to the well-known Rolle theorem from calculus (for example, see [30]), due to the condition (2.7), the optimal interval $[v_1, v_1+\Delta]$ should also always contain the maximum point of the intensity function $I_{BB}(x,T)$. So, the beginning $(v_1)$ and the end $(v_1+\Delta)$ of the optimal interval are located, correspondingly, from the left and right sides of the maximum of the radiant spectral energy density.

$$v_W = 2.8214 \frac{k_B T}{h}. \tag{2.8}$$

The calculated values of the $v_1$, $v_1+\Delta$ (in arbitrary units, correspondingly, $x_1$, $x_1 + s$) are provided in the Table 1. The function

$$R(x_1, x_1 + s) = \frac{E_p(x_1, x_1 + \Delta)}{E_p(0, \infty)} \tag{2.9}$$

is a relative part of the total energy emitted by the interval.

As it can be seen from the Table 1, Wien's maximum $X_{max} = 2.8214$ is always located between the left $x_1$ and right $(x_1 + s)$ ends of the interval. The value $R(x_1, x_1 + s)$ represents a maximum possible value of the relative intensity for a given value of $s$.

| s | $x_1$ | $x_1 + s$ | $R(x_1, x_1 + s)$ | $2\pi A (x_1, x_1 + s)$ |
|---|---|---|---|---|
| 0.1 | 2.7717 | 2.8717 | 0.0219 | 32.37 |
| 0.2 | 2.7225 | 2.9225 | 0.0438 | 16.69 |
| 0.3 | 2.6738 | 2.9738 | 0.0656 | 11.12 |
| 0.5 | 2.5779 | 3.0779 | 0.1091 | 6.65 |
| 1 | 2.3472 | 3.3472 | 0.2161 | 3.33 |
| 1.5 | 2.1294 | 3.6294 | 0.3191 | 2.23 |
| 2.5 | 1.7321 | 4.2321 | 0.5062 | 1.23 |
| 5 | 0.9588 | 5.9588 | 0.8257 | 0.26 |
| 10 | 0.2088 | 10.2088 | 0.9914 | 0.23 |

**Table 1**. Calculated parameters of the Optimized Coherence Filter. The function $A(x_1, x_1 + s)$ is defined in the formula (3.8).

## 3. Coherent properties of the optimal filter

Let us now consider the influence of the proposed optimized filter on the coherence properties of the transmitted thermal light.

To do this, following [1], we consider the superposition of two planar linearly polarized monochromatic waves $E_v(t)$ and $E_v(t+\tau)$ separated by the optical path $c\tau$, where $\tau$ is the time interval, $v$ is the oscillation frequency and $c$ is the speed of light. These waves can be created, for example, by the separation of the same beam using the interferometry methods, or by any other approach. The superposition of these two planar waves leads to the following expression for the resulting intensity

$$I_v(t,\tau) = \langle |E_v(t)|^2 \rangle + \langle |E_v(t+\tau)|^2 \rangle + 2\langle E_v(t)E_v(t+\tau) \rangle. \tag{3.1}$$

Considering the process that is stationary and integrates over all frequencies, it is possible to get the expression for the total energy density of thermal lights passing in both directions through the filter. Under the assumptions made in (2.1) - (2.3) in the previous chapter, the expression for the total energy can be finally represented as:

$$E_P(x_1, x_1+s, \tau) = \frac{1}{2} E_P(x_1, x_1+s, 0)(1+\gamma(x_1, x_1+s, \tau)), \tag{3.2}$$

where the first order temporal self-coherence $\gamma(x_1, x_1+s, \tau)$ can be presented in the form:

$$\gamma(x_1, x_1+s, a) = \frac{\int_{x_1}^{x_2} I_{BB}(x)\cos 2\pi ax\, dx}{\int_{x_1}^{x_2} I_{BB}(x)\, dx}. \tag{3.3}$$

Note that differently from [1], the integrals in (3.3) are taken in a finite interval, due to the used filter. Here $a$ is an arbitrary "time" parameter

$$a = \frac{k_B T \tau}{h} \tag{3.4}$$

So, the degree of the temporal coherence can be characterized as the ratio of energies

$$C(x_1, x_1+s, \tau) = \frac{1}{2}\frac{E_P(x_1, x_1+s, \tau)}{E_P(x_1, x_1+s, 0)} = \frac{1}{2}(1+\gamma(x_1, x_1+s, \tau)), \tag{3.5}$$

where the function $\gamma(x_1, x_1 + s, a)$ is defined by (3.3).

Our results of the calculations of the degree of the temporal coherence function $C(a)$ for optimized filters are presented in Figure 1. The obtained results demonstrate a very strong dependence of $C(a)$ on the values of an arbitrary parameter $s$. By reducing the value of the parameter $s$ it is possible to significantly increase the degree of coherence. Behavior of the function $C(a)$ is different in the following intervals:

a) **$s = \infty$:** in the case of the infinite interval, the calculated function $C(a)$ reproduces the results obtained in [1] and demonstrates very weak level of coherence;

b) **$3.5 < s < \infty$:** function $C(a)$ in this range demonstrates a very weak degree of coherence. The function has only one minimum. Only after reducing the parameter $s$ to 3.5, the level of coherence (or, more exactly, anti-coherence) reaches the significant value of 0.2;

c) **$1.5 < s \leq 3.5$:** in this interval, in addition to the minimum, the function $C(a)$ has also a maximum. When the value of the parameter $s$ is decreased to 1.5, the coherence level at this maximum reaches the significant value of 0.8;

d) **$1.0 < s \leq 1.5$:** in this interval, the function $C(a)$ has 2 minimums and 1 maximum, the second minimum reaches the coherence level of 0.2 at $s = 1.0$;

e) **$0.5 < s \leq 1.0$:** function $C(a)$ has several minimums and maximums in this range;

f) **$0.1 < s \leq 0.5$:** function $C(a)$ quickly oscillates inside of a very slowly changing "envelope";

g) **$0 < s \leq 1.0$:** the changes of "modulating" outer "envelope" are so slow that the function $C(a)$ is almost perfectly harmonizes oscillates over very long time intervals;

h) **$s \rightarrow 0$:** finally, at the limit $s \rightarrow 0$, the filter produces a perfectly coherent thermal light with Wien's maximum frequency for a given temperature:

$$\lim_{s \to 0} v_1 = v_w = \frac{2.8214 k_B T}{h} \quad , \tag{3.6}$$

$$\lim_{s \to 0} C(x_1, x_1 + s, \tau) \equiv \cos 2\pi v_w \tau. \tag{3.7}$$

To quantitatively characterize the degree of coherence for finite values of s, we also introduce an arbitrary coherent "time" A as the maximum value of the parameter a from (3.3), where the value of the degree of coherence function $C(a)$ either exceeds 0.8 or falls below 0.2:

$$A = \max(a); \ (\text{where } C(a) \geq 0.8 \text{ or } C(a) \leq 0.2) \tag{3.8}$$

(i.e. in other words, we consider the significance of coherence, if it either exceeds the threshold of 0.8 for correlation, or 0.2 for anti-correlation).

## 4. Analytical representation of the coherence function.

For a more detailed analysis of the degree of temporal coherence $C(a)$, let's calculate analytically the upper integral in the expression (3.3) for $\gamma(x_1, x_1 + s, a)$. Taking it by parts, we can obtain:

$$\int_{x_1}^{x_1+s} I_{BB}(x) \cos 2\pi ax \, dx = \frac{1}{2\pi a} \left[ I_{BB}(x_1 + s) \sin 2\pi a(x_1 + s) - I_{BB}(x_1) \sin 2\pi a(x_1) - \int_{x_1}^{x_1+s} I'_{BB}(x) \sin 2\pi ax \, dx \right] \tag{4.1}$$

For the optimized bandpass filter, the expression (4.1) can be simplified due to the following reasons:

a) considering the condition (2.7), the values of the intensities $I_{BB}(x)$ at the ends of the intervals are equal to each other;

b) Wien's maximum of $I_{BB}(x)$ is located within the interval, the values of $I'_{BB}(x)$ is equal to zero at Wien's point $X_{max} = 2.8214$, and also take small values around this point. In addition, the values of $I'_{BB}(x)$ have different signs on both sides of $X_{max}$, therefore parts of the integral from $x_1$ to $X_{max}$, and from $X_{max}$ to $x_1 + s$ are partially compensate each other.

So, in the first approximation, the upper integral in (3.3) can be approximated as

$$\int_{x_1}^{x_1+s} I_{BB}(x) \cos 2\pi ax \, dx = \frac{1}{\pi a} I_{BB}(x_1) \sin \pi as \cdot \cos 2\pi as(x_1 + \frac{s}{2}) \ . \tag{4.2}$$

Similarly, the integral in the denominator of the expression (3.3) can be calculated in parts as:

$$\int_{x_1}^{x_1+s} I_{BB}(x)\,dx = I_{BB}(x)x \Big|_{x_1}^{x_1+s} - \int_{x_1}^{x_1+s} I'_{BB}(x)x\,dx = I_{BB}(x_1)s - \int_{x_1}^{x_1+s} I'_{BB}(x)x\,dx \approx I_{BB}(x_1)s. \quad (4.3)$$

Here, we used the same approximations that we used to calculate of the integral (4.2), i.e. considering the closeness to Wien's point $x_w = 2.8214$, $I'_{BB}(x_w) = 0$, where for fairly small values of the interval $s$, we can neglect the second term (which corresponds to a small area above the $I = I_{BB}(x_1)$ line). Finally, in the first order of approximation in the arbitrary parameter $\left(\frac{1}{2\pi a}\right)$, we can get the approximate expression for the function $\gamma(x_1, x_1 + s, a)$:

$$\gamma(x_1, x_1 + s, a) = \frac{1}{\pi as} \sin \pi as \cdot \cos(2\pi a(x_1 + \frac{s}{2})). \quad (4.4)$$

The solution (4.4) is very similar to the expression for the signal after passing the ideal "sinc" filter [31].

As it can be seen from Eq. (4.4.), the solution represents the multiplication of two factors: a) the slowly varying and decreasing "envelope" function and b) high frequency "base" oscillations. The frequency of the "envelope" function is controlled by the width of the interval s. The frequency of the "base" oscillations is given by the central frequency of the interval, which is very close to Wien's frequency).

In particular, for a thermal radiation filter, both the width of the "envelope" and a "base" frequency depend on the temperature of the radiation source (as it follows from the definition of the arbitrary parameters $s$, $x_1$).

Finally, following to (3.5), it leads to the following expression for the degree of temporal coherence $C(x_1, x_1 + s, a)$

$$C(x_1, x_1 + s, a) = \frac{1}{2}\left(1 + \frac{1}{\pi as} \sin \pi as \cdot \cos(2\pi a(x_1 + \frac{s}{2}))\right), \quad (4.5)$$

which can be used with a good accuracy as an interpolation formula for the numerical results obtained in the Chapter 3.

## 5. Conclusion. Possible Applications.

As follows from the data presented in Table 1, even a small change in the $R$ value causes a significant increase in the correlation length. Roughly speaking, by reducing the value of $R$, we can achieve a 3.2 times larger increase in the correlation time. This effect of increasing the optical coherence is a result of the optimization of the filter and can be widely used for various applications.

For example, the maximum total area of single solar antennas or rectangular units that is equal to the second degree of the correlation length (and therefore proportional to $A^2$), can be (as follows from Table 1) doubled if $R$ decreases by 12% or tripled if $R$ decreases by 23%.

Thus, the use of the optimized filters will allow to increase the size of the single rectenna's cells, and make their production much cheaper. Larger cell sizes will also significantly simplify the integration of the multiplier-produced cells.

The obtained results can also be applied to other areas, for example, to increase the width of a layer of multilayer photovoltaic devices based on traditional switching (see Introduction). The thickness of one layer can be increased by 50% in case the total energy is reduced by only 16%.

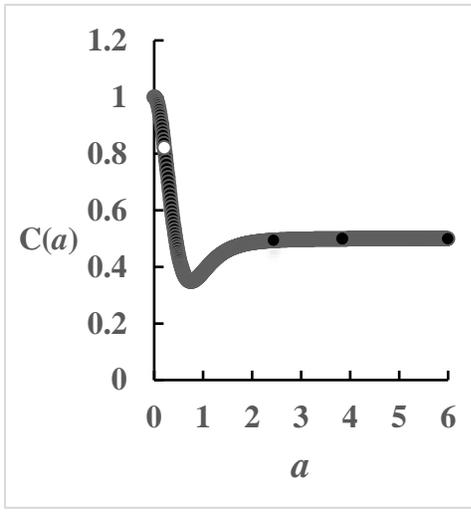 a)

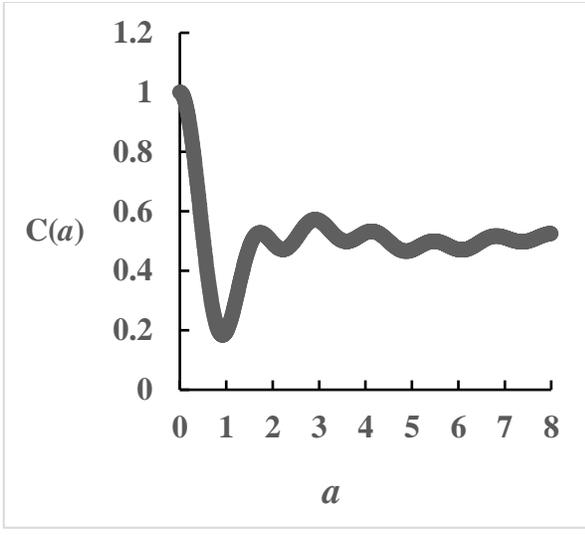 b)

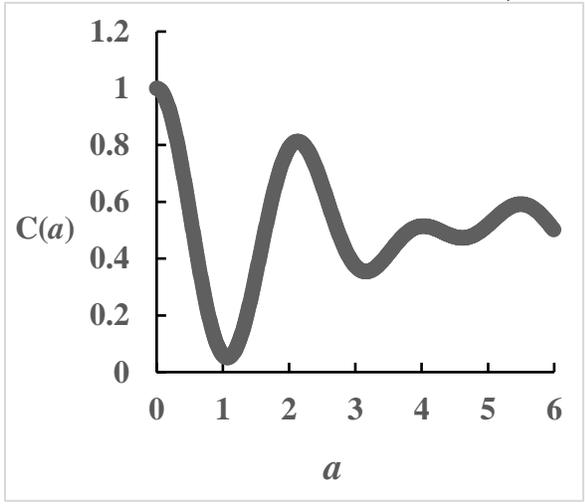 c)

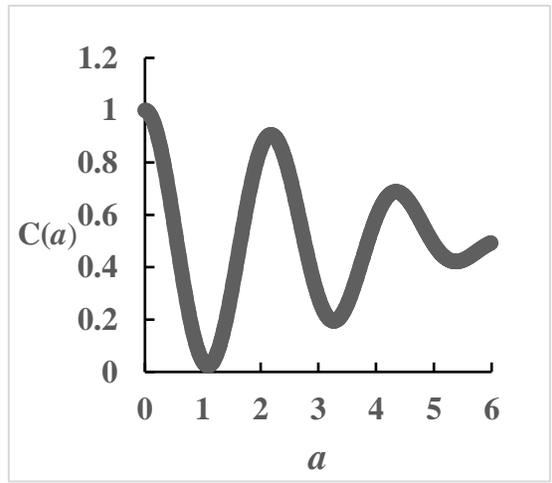 d)

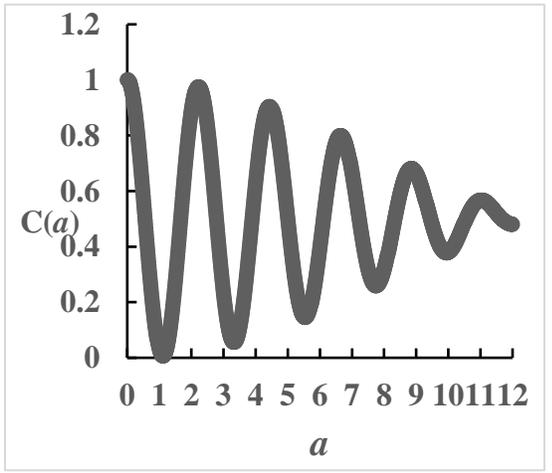 e)

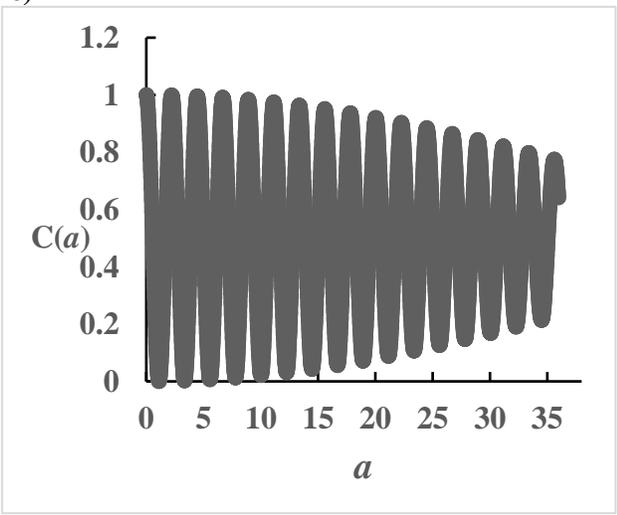 f)

**Figure 1.** The calculated values for the dependences of the degree of the coherence function $C(x_1, x_1 + s, a)$ on the arbitrary time parameter $a$ for the optimized filter for various values of the arbitrary frequency interval $s$.

a) $s = $ infinity; b) $s = 3.5$; c) $s = 1.5$; d) $s = 1.0$; e) $s = 0.5$; f) $s = 0.1$.